\documentclass[11pt,a4paper]{article}
\usepackage{graphicx}
\usepackage{cite}
\usepackage{jpc}
\usepackage{fullpage}
\usepackage{setspace}

\begin{document}
 \doublespacing
 \title{ Relationship between Structure, Entropy and Diffusivity in Water 
 and Water-like Liquids }
 \author{Manish Agarwal$^1$, Murari Singh$^{\dagger}$, Ruchi Sharma$^1$ and
 Mohammad Parvez Alam$^1$\\ and Charusita
Chakravarty$^1$$^*$}

\maketitle
\begin{centering}
	$^1$Department of Chemistry,
	Indian Institute of Technology-Delhi,
	New Delhi: 110016, India.
	$^{\dagger}$School of Physical Sciences, Jawaharlal Nehru University,
	New Delhi: 110067, India
	$^*$Tel:{(+)91-11-2659-1510}
	Fax:{(+)91-11-2686-2122}
	E-mail:{charus@chemistry.iitd.ernet.in}\\

\end{centering}
\clearpage
\newpage

\begin{abstract}
Anomalous behaviour of the excess entropy ($S_e$) and
the associated scaling relationship with diffusivity are compared in 
liquids with very different underlying interactions but similar 
 water-like  anomalies: water (SPC/E and TIP3P models), tetrahedral
 ionic melts (SiO$_2$ and BeF$_2$) and a fluid with core-softened, 
 two-scale ramp (2SRP) interactions. We demonstrate the presence 
of an excess entropy anomaly in the two water models.  
Using length and energy scales appropriate for onset of anomalous
behaviour, the density range of the excess entropy anomaly 
is shown to be much narrower in water than in ionic melts or the 2SRP fluid. 
While the  reduced  diffusivities ($D^*$) conform to the excess 
entropy scaling relation, $D^* =A\exp (\alpha S_e)$ for all the systems 
(Y.~Rosenfeld, Phys. Rev. A {\bf 1977}, {\it 15}, 2545), 
the exponential scaling parameter, $\alpha$, 
shows a small isochore-dependence in the case of water. 
Replacing $S_e$ by pair correlation-based
approximants accentuates the isochore-dependence of the diffusivity scaling.
Isochores with similar diffusivity scaling parameters  are shown to have 
the temperature dependence of the corresponding entropic
contribution. The relationship between diffusivity, excess entropy and pair correlation
approximants to the excess entropy are very similar in all the tetrahedral liquids.
\end{abstract}
\begin{figure*}[h!]
	\begin{center}
		\includegraphics[width=2in,height=2in]{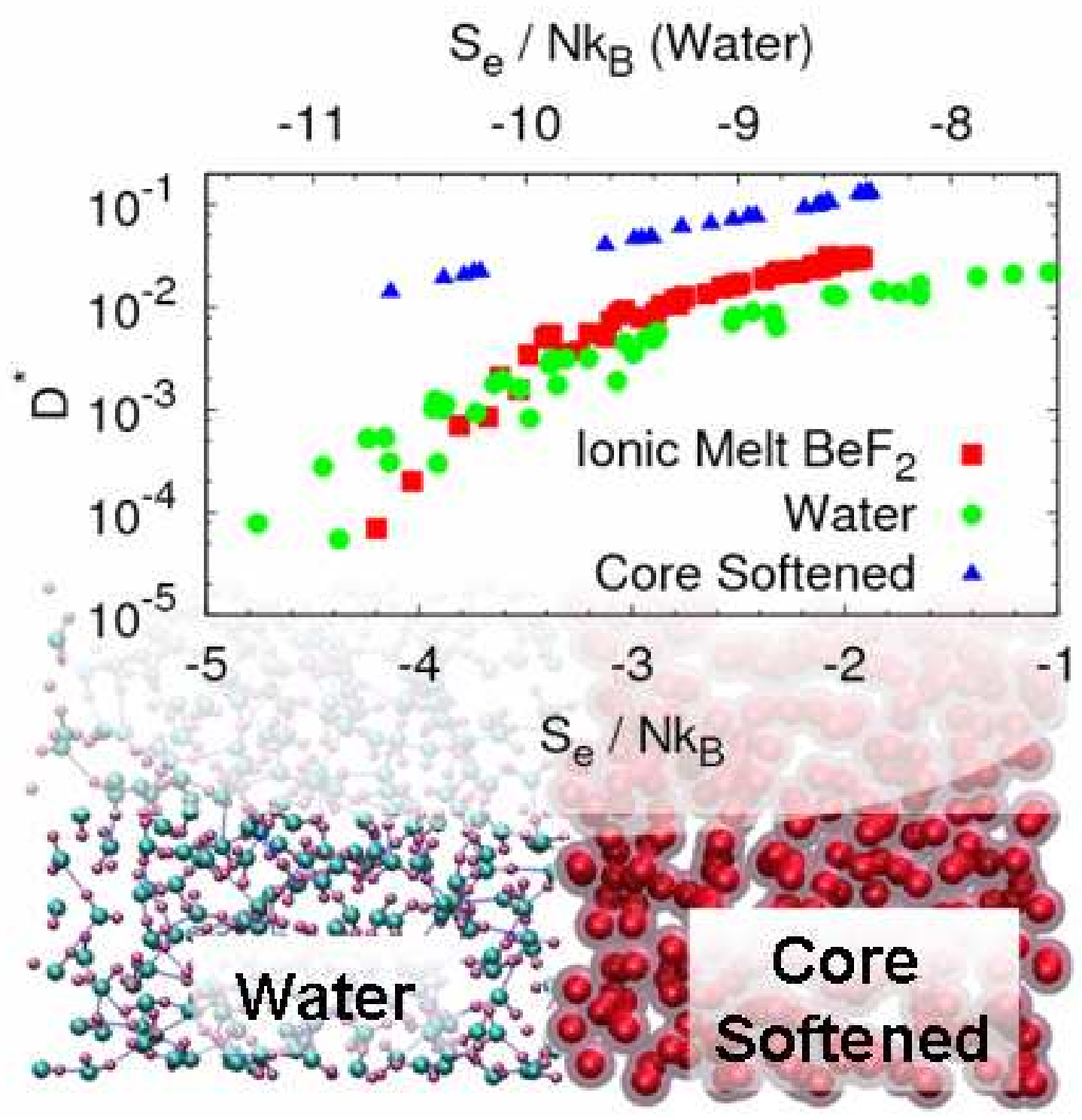}
	\end{center}
\end{figure*}
\clearpage
\newpage

\section{Introduction}

Water displays a number of thermodynamic and kinetic anomalies when compared
to simple liquids \cite{ms98,pgd03,ed01}. 
The density anomaly, corresponding to a negative isobaric thermal
expansion coefficient ($\alpha_P$), is the best known of these unusual 
properties 
of water and is observed for state points lying within an approximately
parabolic boundary defined by the locus of temperatures of maximum density 
(TMD) for which $\alpha_P =0$.
The density anomaly  implies  the presence of other thermodynamic
anomalies, such as those associated with the isobaric 
heat capacity ($C_p$) and the isothermal compressibility ($\kappa_T$).
The kinetic anomalies of water are associated with 
an increase in molecular mobility on isothermal compression, 
measured in terms of diffusivity, orientational relaxation times or viscosity.
A number of network-forming inorganic melts with 
local tetrahedral order have been shown possess water-like anomalies,
most notably, AB$_2$ ionic melts such as SiO$_2$ and BeF$_2$ and 
the liquid phase of elements, such as silicon  and tellurium 
\cite{sslss00,sdp02,pha97,hma01,ht02,scc06,asc07,ac07,acpre09,mm09,sa03,kyy01}.
More recently, water-like anomalies have been demonstrated in mesoscopic
liquids with isotropic, core-softened  effective interactions 
or anisotropic patchy interactions \cite{eaj01,ybgs05,fs08}.  
Despite a very diverse set of underlying interactions,  liquids
with water-like anomalies are found to have essentially conformal liquid-state
``phase diagrams'' with a nested structure of anomalous regimes of density, 
diffusivity and structural order.  

The similarity in the phase diagrams of water-like liquids reflects
similar structure-entropy-diffusivity relationships that can be 
conveniently analysed in terms of the excess entropy, $S_e$, defined as the 
difference between the total thermodynamic entropy ($S$) and the corresponding 
ideal gas entropy ($S_{id}$) at the same temperature and density.
A necessary condition for a fluid to show water-like thermodynamic and
transport anomalies is the existence of an excess entropy anomaly,
corresponding to a rise in excess entropy, $S_e$, on isothermal compression
($(\partial S_e /\partial\rho)_T >0$) 
 \cite{scc06,asc07,ac07,sac08,dfnb08,met061,met10,met062,etm06,ybs08}. 
Liquids with water-like anomalies display distinct forms
of local order or length scales in the low- and high-density regimes;
competition between the two types of local order results
in a rise in excess entropy at intermediate densities.
Most liquids, including anomalous ones, obey semi-quantitative 
excess entropy scaling relationships for transport properties  of the form
$X^*=A\exp (\alpha S_e)$
where $X^*$ are reduced transport coefficients, and $A$ and $\alpha$
are scaling parameters that are very similar for systems with conformal
potentials \cite{yr771,yr99,md96,has00}. Consequently, the existence of an
excess entropy anomaly is reflected in mobility anomalies. 
In order to make a more precise  connection between thermodynamic
and mobility anomalies, it is necessary to understand  
the assumptions underlying the scaling rule.
Rosenfeld-scaling assumes  that diffusion
in liquids  takes place through a combination of binary collisions 
and cage relaxation. The binary collision  contribution is approximately
factored out by using macroscopic reduction parameters based on 
elementary kinetic theory. For example, reduced diffusivities are defined
as $D^* = D(\rho^{1/3}/(mk_BT)^{1/2})$ \cite{yr99}.
The frequency of cage relaxations is assumed to be
proportional to the number of accessible
configurations, $\exp(\alpha S_e)$, since configuration space connectivity
is high in the stable liquid phase. In order for the scaling relationship
to be state-point independent, it is necessary that
the exponential parameter $\alpha$ controlling the number of accessible
configurations is determined by the interaction potential and is
otherwise state-point independent.

This paper develops a basis for quantitative comparison of
structure-entropy-diffusivity relationships in liquids with very different
underlying interactions but similar water-like anomalies.
We focus on three different categories of liquids: (i) molecular fluids
(H$_2$O), (ii) tetrahedral ionic melts (BeF$_2$ and SiO$_2$) and
a core-softened fluid with isotropic, pair-additive interactions.
Comparison of these  different systems should provide very useful physical 
insights into the relationship between
structure, entropy and transport properties of liquids, including
complex fluids, and has not been previously attempted. 
It is evident that parametric variations
of a single functional form for the interaction potential are
insufficient to serve as a basis for comparison of anomalous behaviour for
the three different categories of liquids studied here. Instead, we focus on
directly comparing the two crucial features of the excess entropy relevant
for water-like liquids: firstly, the presence of an excess entropy anomaly and
secondly, exponential scaling of transport properties with excess entropy.
Since the state point, ($\rho_m, T_m$), corresponding to the maximum 
temperature along the TMD locus is a unique point common to all liquids
with a water-like density anomaly, we use this information to define a
natural energy and length scale for onset of anomalous behaviour 
in each system. 

The specific interaction models used for the different categories of liquids
require some introductory comments. Anomalous behaviour and scaling of
transport properties with the thermodynamic excess entropy has so far
not been studied in the case of water. Since the different water models
are known to display anomalous properties at very different temperatures,
we use two different effective pair potentials for water, TIP3P and
SPC/E. The SPC/E model is known to qualitatively reproduce all anomalous
properties of water but typically at temperatures that are 30K to 40K
lower than the experimental value of 279K\cite{bgs87,ed01,pgd03,sslss00,sac08}. The TIP3P model is widely used in biomolecular simulations, but 
the onset temperature for anomalous behaviour is 170-180K
\cite{jcmik83,sac08,va05}. The transferable rigid-ion model (TRIM) potential was used for BeF$_2$\cite{wac76} and  and the
van Beest-Kramer-van Santen (BKS) potential was used for SiO$_2$ \cite{bks90}. As an
example of a core-softened fluid with waterlike anomalies, we consider
the two-scale ramp potential (2SRP) \cite{eaj01,ybgs05}. Thus we study five different simulation
models for the three categories of liquids.
Table 1 summarises some of the relevant information about these simulation
systems at 
$(\rho_m, T_m)$. Section 2 provides the necessary simulation details.
Results are presented in Section 3 and conclusions are discussed in section 4.

\section{Simulation Details}

Potential energy models for all systems studied here are given
in Table 1. Molecular dynamics simulations of ionic melts (BeF$_2$ and SiO$_2$) and
water models (TIP3P and SPC/E) were performed
in the NVT ensemble using the Verlet algorithm as implemented
in the DL\_POLY package\cite{syr02}. Temperature range of 1500-3000K
is studied for BeF$_2$ and 4000-6000K for SiO$_2$.
Details of simulations of the two
ionic melts are given in ref.\cite{agc09}. Simulations of TIP3P and SPC/E
water used 256 molecules in a cubic simulation box with 1 fs timestep and production run lengths of
4 to 8 ns. Rigid-body constraints were maintained using the SHAKE algorithm.
The Berendsen thermostat time constant, $\tau_B$, is  1ps for
SPC/E for all state points except those along the 210K isotherm ($\tau_B$=200ps)
while $\tau_B$= 20ps 
for all TIP3P state points. State points for SPC/E water covered a 
temperature range from 210K to 300K and density range from 0.9 g cm$^{-3}$ 
to 1.4 g cm$^{-3}$; the temperature range for TIP3P water was kept as
170K to 300K.

Simulations of the two-scale ramp potential (2SRP) fluid were
performed using Metropolis Monte Carlo simulations using 256 particles
in a cubic simulation box. The 2SRP fluid has length scales
associated with the hard-core ($\sigma_0$) and soft-core ($\sigma_1$)
diameters; we use $\sigma_1$ as the unit of length. The energy $U_1$ of the 
linear ramp interaction extrapolated to zero separation is taken
as the unit of energy. State points cover a temperature range of 0.025
to 0.2 and  density range from 0.8 to 2.6 in reduced units 
\cite{scc06}). Diffusivities for five isotherms from $T=0.027$ to $T=0.063$ 
were taken from ref.\cite{kbszs05}.

 Widom insertion method was used to calculate 
chemical potential and  excess entropy for the 2-scale ramp fluid
at low densities and
high temperatures \cite{fs02}. Thermodynamic integration was then used to
obtain excess entropies at other statepoints. For ionic melts, $S_{id}$ is
computed for a non-interacting multi-component mixture of atoms with the same
masses and mole fractions as the ionic melt. Excess entropy for BeF$_2$ and
SiO$_2$ melts were calculated using the same procedure as in ref\cite{sac08},
but over a more extensive density range. The excess entropy of the  water
models was computed relative to an ideal gas of rigid triatomic molecules
at 1000K, 0.01g/cc. 
Thermodynamic integration was used to
compute $S_e$ at other state points \cite{sslss00}. Our entropies at
298K and 1 g cm$^{-3}$ match those of \cite{rhh07} for SPC/E and TIP3P, within
statistical error.
The use	of atom-atom RDFs in equation(1) would suggest that an ideal
		gas reference state analogous to that used for ionic melts would
		be more appropriate; this is in fact not necessary, since the
		difference between the two reference states will be constant
		at a given temperature and density. We note that while the
system sizes chosen were chosen to be small enough that a large number of state
points could be efficiently covered, they are sufficiently big   that 
finite-size effects would
be expected to make only small quantitative differences to the computed values
of thermodynamic and transport properties.

\section{Results}

\subsection{Excess Entropy Anomaly}
 The strength of the excess entropy anomaly  determines the 
 presence of diffusional, density and related anomalies and  reflects
 the variation in  structural order in a fluid as a function of density.
  Figure 1  compares the density dependence of the excess 
 entropy
 along the $T=T_m$ isotherm, scaling densities with respect to the
 $\rho_m$ value of each system.
While the overall behaviour of $S_e(\rho /\rho_m)$ is
very similar, it is immediately
apparent that, in units of $\rho_m$, the water models have a relatively narrow
density regime of anomalous behaviour and a steeper slope in the high density regime.
The two-scale ramp (2SRP) system in contrast, has a much wider anomalous
regime bounded by a  well-defined minimum as well as maximum in $S_e(\rho) $ 
along the $T_m$ isotherm. The two ionic melts resemble the ramp fluid
in the density range of anomalous behaviour and the location of the maximum in
$S_e (\rho /\rho_m)$. The minimum in the $S_e(\rho /\rho_m)$
curves  could not, however, be located within the density range consistent with
the thermodynamic stability criterion that the isothermal compressibility should
be positive. In units of $Nk_B$, the rise in the $S_e(\rho )$ curves from
the minimum to the maximum is a fraction of a $k_B$ for all the systems.
This small anomaly is, however, sufficient to create an inflexion point in the
total $S(\rho ) =S_{id} + S_e$ curves that marks the onset of the density
anomaly.  
The qualitative behaviour of the excess entropy anomaly is consistent with the 
grouping of the fluids into three similar classes: rigid water models,
ionic melts and core-softened fluids. Unless otherwise stated, results for
only one member of each class will be discussed in the rest of this paper.

\subsection{Excess Entropy Scaling of Diffusivities}

A comparison of the excess entropy scaling of the diffusivities for
the different systems studied in this work, shown in Figure 2, illustrates that
such a scaling  approach is  a  reasonable predictor 
of diffusivities though there are small system-dependent differences. 
The 2SRP fluid shows excellent scaling  for the
state points studied, but the range of variation of the
diffusivities is small in comparison to  both ionic melts
and water, given temperature variations by factors of two to three.
Both SiO$_2$ and BeF$_2$ show very good scaling  
in the liquid regime, but there is a change in slope $\alpha$ for 
the lowest temperature isotherms, possibly due to 
onset of cooperative dynamics \cite{md99,khd05,met061,met10,agc09}.
The water models all show a distinct isochore dependence of
the $\ln D^*$ versus $S_e$ plots, which is less pronounced in ionic melts
and is essentially zero in the 2SRP fluid.
In the case of water, deviations from linearity are specially noticeable at 
low temperatures
for the 0.90 and 1.40 g cm$^{-3}$ isochores for which the mode-coupling
temperature is relatively high \cite{sss99}. All liquids with water-like anomalies
are associated with a competition between two different length scales 
or local order metrics as a function of density. Therefore deviations
from corresponding states behaviour would be expected. It appears, however,
that these deviations are small and most noticeable in the case of water
which has a rigid, non-spherical structure. Possible reasons for this are discussed
below. 

\subsection{Pair Correlation Estimators of the Excess Entropy}

The effect of the underlying structural correlations in
a fluid  on the entropy-transport relationships can be understood
by expressing $S_e$ as a multiparticle correlation expansion,
$S_e=S_2 + S_3 + \dots$, where $S_n$
is the entropy contribution due to $n$-particle spatial correlations
\cite{hsg52,be89,lh92}. The entropy contribution 
due to pair correlations between atoms of type $\alpha$ and $\beta$ is given by:
\begin{equation}
S_{\alpha\beta} = \int_0^\infty 
\{g_{\alpha\beta}(r)\ln g_{\alpha\beta}(r) -\left[ g_{\alpha\beta}(r)-1\right] \} r^2 dr
	\label{s2calc}
\end{equation}
where $g_{\alpha\beta}(r)$ is the atom-atom pair distribution function (PDF).
The overall pair correlation  entropy, $S_2$, is:
$S_2/Nk_B = -2\pi\rho\sum_{\alpha ,\beta } x_{\alpha }x_{\beta }S_{\alpha\beta }$
where $N$ is the number of particles and  $x_\alpha$ is the mole fraction of 
component $\alpha$ in the mixture.  $S_2$ can be estimated
from experimental scattering data as well as simulations and  is
the dominant contribution to the excess entropy in the case of simple\cite{be89} and 
core-softened fluids (e.g. results for 2SRP fluid in this paper), ionic melts and water \cite{sac08}. 
 In the case of the AB$_2$ melts and water, we also compute  the tetrahedral correlation entropy, $S_{AA}$, due to the pair correlations  between the
tetrahedral(A) atoms, corresponding to Be, Si and O in BeF$_2$, SiO$_2$ and 
H$_2$O respectively.  The $S_{OO}$ entropic contribution is reproduced 
by many coarse-grained model potentials of 
water that replace each water molecule by a structureless particle
with isotropic effective  interactions that  reproduce the 
behaviour of $g_{OO}(r)$ \cite{jh09,cs09}. 

\subsubsection{Diffusivity Scaling}

The effect of using  a structure-based approximant to the excess
entropy, such as $S_2$, on the Rosenfeld-scaling relation can be seen
by substituting  $S_e =S_2 + \Delta S$  to obtain
\begin{equation}
	D^*\approx A\exp (\alpha S_e) =A\exp\left[\alpha S_2(1+ \Delta
	S/S_2)\right]
	=A\exp(\alpha 'S_2)
\end{equation}
where the new exponential scaling parameter, $\alpha ' =1 + \Delta S/S_2$,
has a state-point dependence that is determined by the density and
temperature-dependent behaviour of $\Delta S/S_2$. Thus by using
different approximants to the excess entropy and observing the
effect on the scaling behaviour of the transport properties, one
can determine importance of different structural correlations
in controlling mobility.
Figure 3 shows the correlation between the dimensionless diffusivity ($D^*$)
and $S_2$ as well as $x_{OO}^2S_{OO}$ for SPC/E water, which should be
compared with Figure 2(a). It is immediately obvious that the temperature
dependence along an isochore is linear or quasi-linear, but there is
a distinct isochore dependence of the slope $\alpha$ of the $\ln D^*$ versus
$S_\mu$ plots. The isochore-dependence is most pronounced for $S_\mu =S_{OO}$
and least pronounced for $S_\mu =S_e$. While the
$\ln D^*$ versus $S_2$ or $S_{AA}$ behaviour has been studied before 
\cite{met062,ybs08,acpre09,agc09}, no systematic comparison with $S_e$ and 
$S_2$ scaling has been made so far.
The ionic melts show a parallel behaviour to the water models, in terms
of diffusivity scaling with respect to $S_e$, $S_2$ and $S_{AA}$ while the
2SRP fluid also shows a small isochore dependence when $D^*$ is scaled with
respect to $S_2$. In the case of BeF$_2$, the diffusivity and viscosity
scaling with respect to $S_e$ and $S_2$ is given in refs.\cite{acpre09,agc09} and
with respect to $S_{BeBe}$ in  Supporting Information Figures 1. Scaling with respect
to both $S_e$ and $S_2$ for 2SRP fluid is shown in Supporting Information Figure 2.
This is in contrast to  simple liquids with pair additive potentials for
where excess entropy
scaling relationships based on $S_e$ and $S_2$ have a similar degree of 
universality \cite{md96}.

\subsubsection{Temperature Dependence}

The isochore-dependence of diffusivity scaling with different entropic
contributions, and the contrast with simple
liquids, suggests that the isochoric temperature
dependence of different entropic measures determines
the quality of entropy-diffusivity scaling.
Based on measure-free density functional theory for repulsive potentials, 
Rosenfeld predicted a  $T^{-2/5}$ scaling for the
excess entropy \cite{rt98}. In the case of simple liquids, these predictions are satisfied
by both $S_e$ and $S_2$ and Rosenfeld-scaling holds very well \cite{rt98,ccpre07}.
$S_e$ is known to show  $T^{-2/5}$ scaling in ionic melts and water
over  temperature regime of interest in this study \cite{sslss00,pha97}.  
Figure 4 shows $S_e$, $S_2$ and $S_{OO}$
as a function of $T^{-2/5}$ for different isochores in SPC/E water. 
Unlike in simple liquids, $S_2$ and $S_{OO}$ show significant deviations from
$T^{-2/5}$ scaling in the anomalous regime.
The results for BeF$_2$  are very similar to
those of SPC/E water (see Supporting Information Figure 3 and ref.\cite{asc07,sac08}). 
In the case of the 2SRP fluid, both $S_e$ and $S_2$ show  deviations from $T^{-2/5}$ behaviour at low temperatures  (see Supporting Information Figure 4).

A comparison of the isochore-dependence of the
Rosenfeld-scaling parameter with respect to an entropic
contribution, $S_\mu$, and the temperature dependence of
$S_{\mu}$ shows that isochores with very different temperature
dependences also have very different diffusivity scaling
parameters.  For example,
Figure 3(b) shows when $\ln D^*$ is plotted as a function of $S_{OO}$, then
the slope increases with increasing density and the three densest isochores,
at $\rho = 1.2$, 1.3 and 1.4  g~cm$^{-3}$ show very different and
progressively increasing values of $\alpha$. Figure 4(c) shows that $S_{OO}$
along these three isochores has a distinctly different and  very weak 
temperature dependence.  The two AB$_2$ ionic melts 
show a very similar pattern of behaviour (see Supporting Information
Figures 1 and 3),
reflecting the similar structure of the tetrahedral fluids.
The 2SRP fluid shows a weak isochore dependence of $\alpha$ when 
$S_e$ is replaced by $S_2$, which may be related to a wider spread of
$S_2$ values compared with $S_e$. It is interesting to note that $S_2(T)$ and
$S_{AA}(T)$ curves show a much stronger density dependence than $S_e(T)$.
In the case of the tetrahedral liquids, it is notable that at high
densities, $S_{AA}$ remains almost constant with temperature while
diffusivity rises quite sharply. Since the overall excess entropy
scaling for high and low density isotherms is not qualitatively different,
this must imply that significant reorganisation of the network associated
with the other pair correlation contributions contributes to controlling
diffusivity; for example, in water this could be strong
librational-translational coupling reflected in the $S_{OH}$ and $S_{HH}$
contributions.

\section{Discussion and Conclusions}

The excess entropy anomaly is compared in five liquids with water-like anomalies, with three
completely different types of interactions: ionic melts (BeF$_2$ and SiO$_2$), hydrogen-bonded molecular liquid (SPC/E and TIP3P water), core-softened fluid (two-scale ramp).
Using  the state point, ($\rho_m, T_m$), corresponding to the maximum 
temperature along the TMD locus to define energy and length scales, we show
that excess entropy anomaly in the ionic melts and the core-softened fluid
is very similar in range and strength, while the two water models have 
a conspicuously smaller density range and a much sharper decrease in $S_e$ with density
on strong compression. 

We show that scaling of
diffusivities with the thermodynamic excess entropy is good in all cases,
noting  that water is the first molecular fluid, other than the Lennard
-Jones chain fluid\cite{gpmc08}, for which Rosenfeld-scaling with the
thermodynamic excess entropy, rather than pair-correlation based estimators,
has been tested.  In comparison to ionic melts and the 2SRP fluid, a  small  
isochoric dependence of the Rosenfeld-scaling parameters is seen in the case of water.
This suggests the presence of  additional length scales due to the   rigid,
non-spherical molecular shape, but these are not critical enough
to destroy the overall correlation. The results for water suggest that Rosenfeld-scaling
will be useful for many other molecular fluids as a predictor of dynamical
properties.

We examine the diffusivity scaling with two different entropy estimators
based on pair correlation distribution functions (PDFs). The first one
which we refer to as the pair correlation entropy, $S_2$, includes information
from all atom-atom PDFs. The second one, referred to as the tetrahedral
correlation entropy ($S_{AA}$),  considers only the contribution from
the pair correlations between the tetrahedral atom-atom PDFs and is applicable
to only water and ionic melts. We show that the isochore dependence of the
diffusivity scaling parameters is most pronounced for $S_{AA}$, much less
pronounced for $S_2$ and virtually negligible for $S_e$.
Thus the state-dependence of Rosenfeld scaling parameters increases as the structural
approximants to the entropy become poorer. We demonstrate that isochores with
similar temperature-dependence of the entropy estimators show very similar 
Rosenfeld-scaling parameters. 

The clear correspondence between the isochoric Rosenfeld-scaling
parameters and the temperature dependence of the entropy along isochores has an
interesting implication. The isochoric temperature dependence of the excess entropy
of many atomic systems obeys a $T^{-2/5}$ scaling,  possibly because of the
critical role of strong, short-range repulsions \cite{rt98}. $S_2$ shows a similar
$T^{-2/5}$ dependence in simple liquids but not in anomalous fluids, as shown
in this study and elsewhere \cite{ccpre07,ac07,sac08}. This suggests that 
diffusivities for many atomic fluids will obey Rosenfeld-scaling best when the
the thermodynamic excess entropy is used. The 2SRP fluid presents an interesting
case in that Rosenfeld-scaling is applicable though there are deviations from
$T^{-2/5}$ scaling. It would be interesting to examine these ideas in the
context of recent work on strongly correlating fluids \cite{pbsd08,gspbd09}.

Our results on structure-entropy-diffusivity relationships are experimentally
testable.  Atom-atom PDFs,  diffusivities
and calorimetric  entropy  are experimentally accessible
as a function of temperature and density (or pressure). A comparison of diffusivity scaling
with the thermodynamic entropy as well as the O-O pair correlation entropy between
water and methanol should itself be very interesting, because they are both molecular, hydrogen-bonded liquids. Methanol, unlike water, does not show pronounced liquid state anomalies and we expect that this will show up in the diffusivity scaling with respect 
to the oxygen-oxygen pair correlation entropy but not with respect to the
thermodynamic excess entropy.
Since the atom-atom RDFs can be measured for
essentially all molecular and polymeric fluids, our results suggest that
excess entropy scaling should have substantial predictive value when combined
with structural data, provided the complete set of atom-atom RDFs is used
for estimating the entropy. However, coarse-graining procedures
which rely on mapping a subset of atom-atom PDFs will  not
necessarily preserve the thermodynamic state-point dependence 
of transport properties, specially in anomalous fluids, as has been
noted in recent studies of coarse-grained potentials for water \cite{jh09,cs09}.

Our results suggest that Rosenfeld-scaling is valid for a wide
range of simple and anomalous liquids.The deviations are small if the complete thermodynamic excess entropy is used, possibly because of the dominance of strong, short-range repulsions. A microscopic understanding of Rosenfeld-scaling is,
however, necessary in order to understand  the occurrence and magnitude
of deviations in different systems. One possible route to developing
such a microscopic approach is to use energy landscape approaches
that have proved very useful in the context of supercooled liquids \cite{ds01,kds09}.
Efforts to apply energy landscape analysis to understand Rosenfeld-scaling
in simple and anomalous liquids have been relatively few \cite{jh09,cc06,oscb10}.

Water-like liquids are systems where an additional length scale comes into
play as a function of density and therefore deviations from Rosenfeld-scaling
may be expected. As discussed in this paper, deviations are, in fact, small in this
case. An analogous set of systems could be those where an additional length scale
emerges as a function of temperature, as in inverse melting systems \cite{fds03}. 
Some of the recent work on Rosenfeld-scaling in systems with finite repulsions,
such as the Hertzian sphere and the Gaussian-core \cite{kkmet09,pcf09,fgr09,kpgset09},
can be understood in terms of deviations from Rosenfeld-scaling behaviour
due to change in the character of binary collisions as well as the possibility of
additional length scales as a function of density.

To summarize, the anomalous behaviour of the excess entropy ($S_e$) and
the associated scaling relationship with diffusivity is compared in five liquids
with water-like thermodynamic and transport anomalies: SPC/E and TIP3P
water models, SiO$_2$ and BeF$_2$ ionic melts and the two-scale ramp (2SRP)
fluid. By defining suitable length and energy scales for onset of anomalous
behaviour, the density range of the excess entropy anomaly in water
is shown to be much narrower in water than in ionic melts or the 2SRP fluid. While the 
Rosenfeld prediction of an exponential dependence  of 
reduced  diffusivity ($D^*$) on the excess entropy ($S_e$), 
$D^* =A\exp (\alpha S_e)$ is found to be be almost quantitatively valid for
all the systems, the exponential scaling parameter, $\alpha$, is found to
show a small isochore-dependence in the case of water, presumably due to
the rigid molecular structure. Unlike in the case of simple
liquids, this isochore-dependence of the excess entropy
scaling of the diffusivities is accentuated by replacing $S_e$ by the
pair correlation entropy, $S_2$, for all the systems. In the case of the
AB$_2$ ionic melts and water, the effect is more pronounced if
the diffusivity is scaled with respect to the pair correlation entropy
associated with the tetrahedral atoms. More interestingly, variations in the
the isochoric temperature dependence of the entropic contribution
($S_e$, $S_2$ and $S_{AA}$) are found to be correlated with variations
in the isochore diffusivity scaling parameters, suggesting that
the organisation of the energy landscape changes significantly with density.
The tetrahedral liquids (ionic melts and water) show very similar
structure-entropy-diffusivity relations.
Our results suggest that by systematically considering systems which represent
different types and degrees of deviation from assumptions underlying
Rosenfeld-scaling, interesting physical insights into the relationship
between structure, thermodynamics and transport in liquids can be obtained.

{\bf Acknowledgements} 
This work was financially supported by the Department of Science and
Technology, New Delhi.
MA would like to thank Indian Institute of Technology-Delhi
for award of a Senior Research Fellowship.
MS would like to thank University Grants Commission, New Delhi for award of a 
Junior Research Fellowship. The authors thank Divya Nayar for assistance with
some of the computations.

{\bf Supporting Information Available:} Additional figures, referenced in
the text to support our conclusions, 
are available free of charge via the Internet at http://pubs.acs.org.


\setcounter{figure}{0}
\clearpage
\section*{Figures}

\begin{enumerate}
	\item Plot of excess entropy $S_e$ with scaled number density,
	$\rho/\rho_m$ for (a) ionic melts and 2SRP fluid and (b) water models. The
	lines mark the respective minima and maxima in the 2SRP fluid in part (a)
	and the SPC/E model in part (b).
\item Correlation plots of Rosenfeld-scaled diffusivities with
	excess entropy, $S_e$ of (a) two-scale ramp potential (2SRP) fluid,
	(b) AB$_2$ Ionic Melts (BeF$_2$ and SiO$_2$) and (c) SPC/E and (d) TIP3P. 
	The lowest isotherms for BeF$_2$ and SiO$_2$
	are shown with filled symbols. Data points lying along the highest and lowest isochore for (c)
	and (d) are joined with smooth lines. The scaling parameter $\alpha$ for
	each plot is shown. In part (b), the datapoints at 1500K(BeF$_2$) and
	4000K(SiO$_2$) are shown as filled symbols, and are excluded from the
	fit. In part (c) and (d), data points having $S_e <$-9.5 are
	excluded from the fit.
\item Correlation plots of Rosenfeld scaled diffusivities, $D^*$,  
	in SPC/E water with: (a)$S_{2}$ (b)$S_{OO}$ for selected isochores.
	Straight lines connect data from the highest and lowest density
	isochores.
\item Different entropic contributions in SPC/E as a function of $T^{-2/5}$:
	(a) thermodynamic excess
	entropy, $S_e$, (b) pair correlation entropy, $S_2$ and (c) entropic
	contribution due to oxygen-oxygen pair correlations, $x_o^2 S_{OO}$ with
	$T^{-2/5}$.
\end{enumerate}

\clearpage
\begin{figure}[h]
	\begin{center}
		\includegraphics[clip=true,trim=0 8cm 0 8cm]{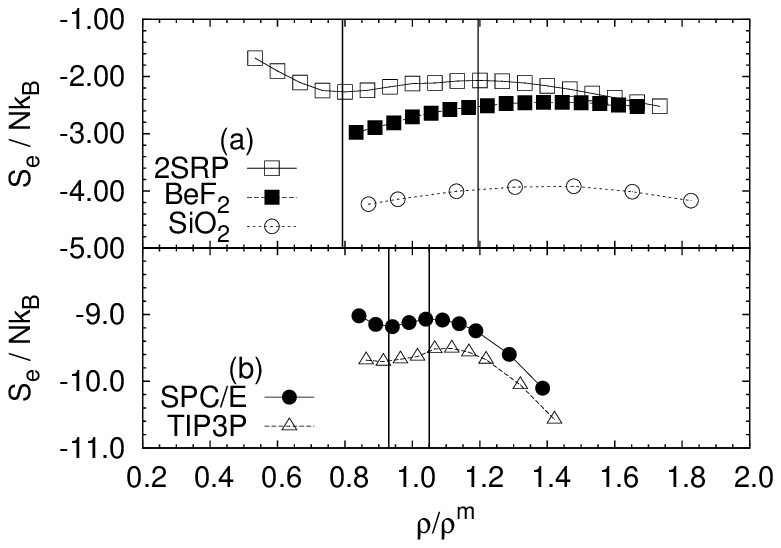}
	\end{center}
	\caption{\label{fig:sesplit}}
\end{figure}
\clearpage

\begin{figure}[h]
	\begin{center}
	\includegraphics[clip=true,trim=0 4cm 0 4cm]{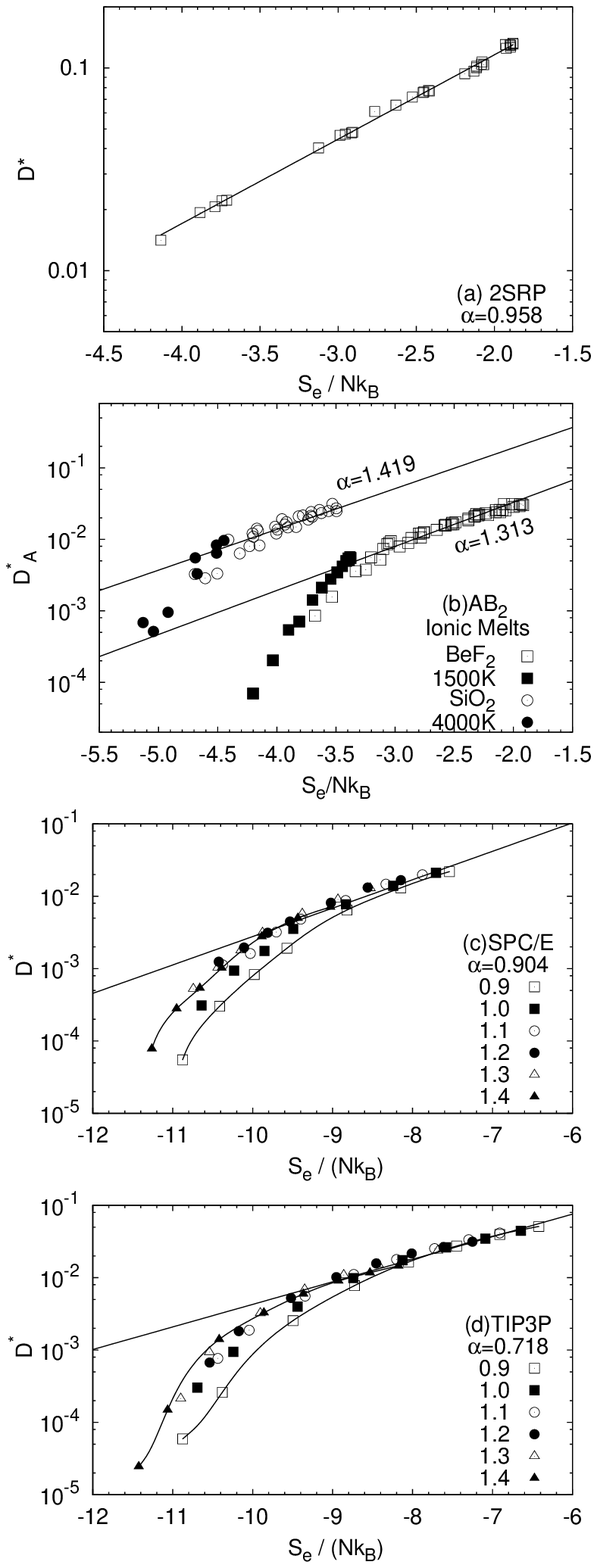}
	\end{center}
	\caption{\label{fig:rosenall}}
\end{figure}
\clearpage

\begin{figure}[h]
	\begin{center}
		\includegraphics[clip=true,trim=0 8cm 0 8cm]{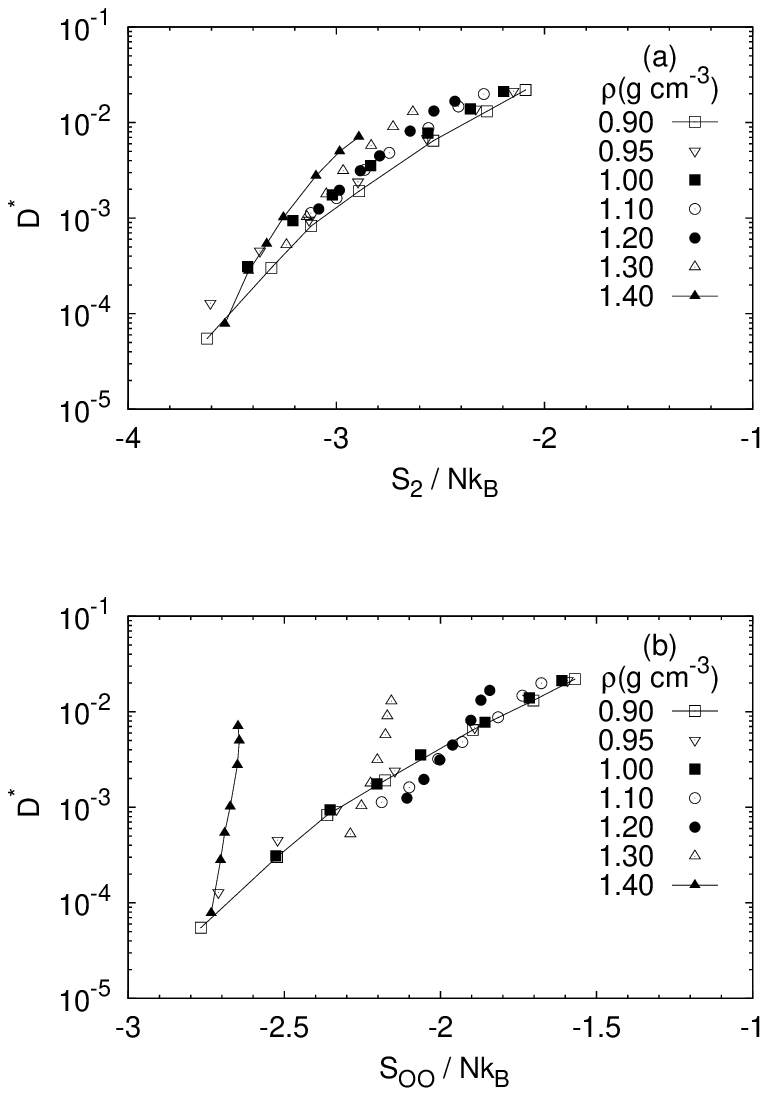}
	\end{center}
	\caption{
	\label{fig:spce3fig}}
\end{figure}
\clearpage

\begin{figure}[h]
	\begin{center}
		\includegraphics[clip=true,trim=0 5cm 0 5cm]{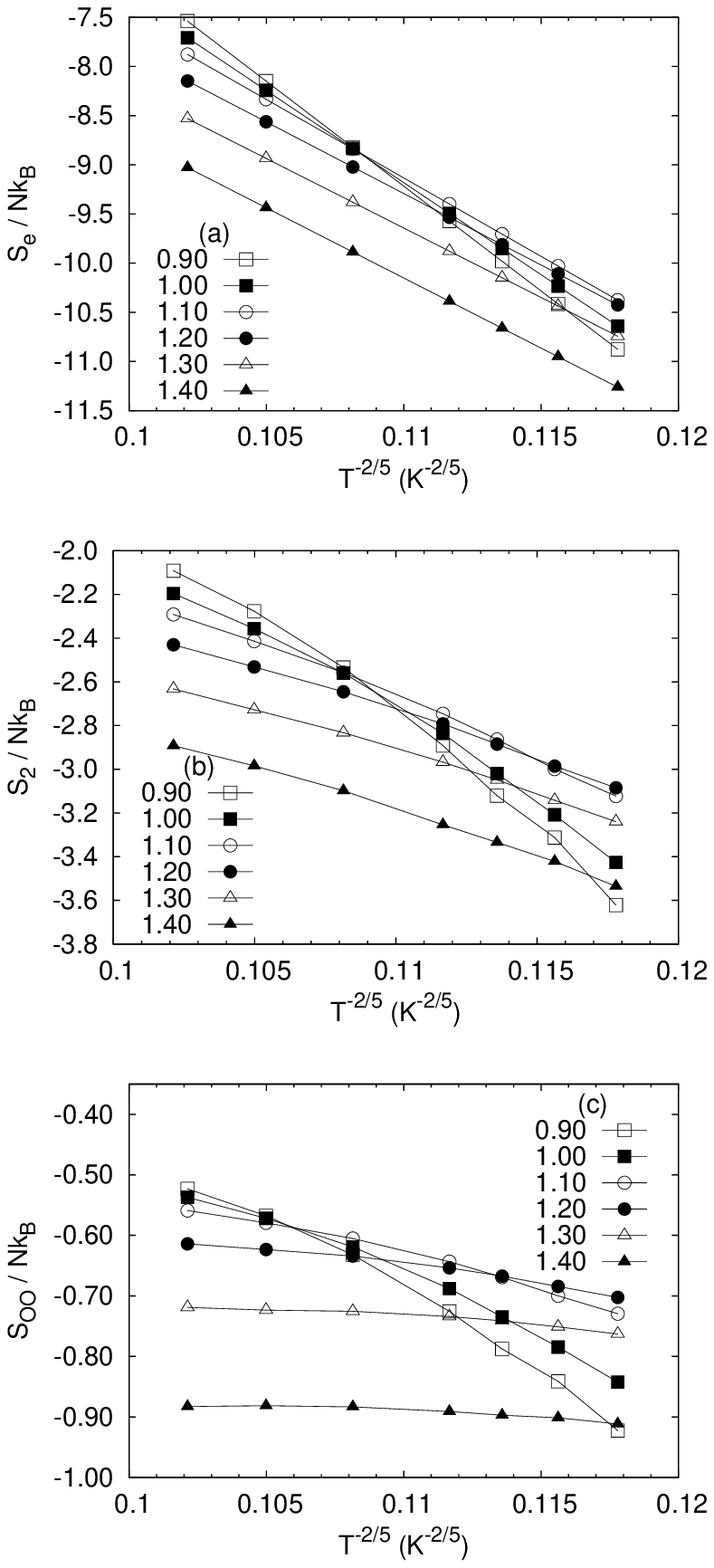}
	\end{center}
	\caption{ \label{fig:se-T}}
\end{figure}
\clearpage

\begin{table}
	\centering
	\begin{tabular}{c|cc|cc|c}
		\hline
		\hline
		~ & SPC/E\cite{bgs87} & TIP3P \cite{jcmik83} &
		BeF$_2$\cite{wac76} & SiO$_2$\cite{bks90} & 2SRP\cite{eaj01} \\
		\hline
		$\rho^m$(g cm$^{-3}$) & 1.01 & 0.98 & 1.8 & 2.3 & 1.5 \\
		$T^m$(K) & 251 & 195 & 2310 & 5000 & 0.0548\\
		\hline
	\end{tabular}
	\caption{State point corresponding to the maximum temperature along the
	TMD locus ($\rho^m$, $T^m$) for the systems studied. Potential
	parameters are given in the references. Values for the 2SRP are
	in the reduced units defined in the simulations details. }
	\label{tab:tmdmaxdata}
\end{table}

\end{document}